\RequirePackage{fix-cm}
\documentclass[twocolumn]{svjour3}
\usepackage[T1]{fontenc}
\usepackage[utf8]{inputenc}
\usepackage{hyperref}
\usepackage{booktabs}					
\usepackage{rotating}
\usepackage{multirow}					
\usepackage{colortbl}					
\usepackage{graphicx}
\usepackage{subcaption}
\usepackage{amsmath,amssymb,amsfonts}
\usepackage{algorithm}
\usepackage{algpseudocode}
\usepackage{comment}
\usepackage{color}
\usepackage{pifont}
\usepackage{xifthen}
\usepackage[table,xcdraw]{xcolor}
\usepackage{enumitem}
\usepackage{bm}
\usepackage{array}
\usepackage{hhline}

\newcommand{\ie}{\emph{i.e.}}
\newcommand{\eg}{\emph{e.g.}}
\newcommand{\etc}{\emph{etc.}}
\newcommand{\ea}{\emph{et al.}}
\newcommand{\vs}{\emph{vs.}}

\newcolumntype{C}{>{\centering\arraybackslash}p{2em}}

\DeclareGraphicsExtensions{.pdf,.png,.jpg}

\journalname{International Journal of Data Science and Analytics}

\begin{document}

\title{You Must Have Clicked on this Ad by Mistake!}
\subtitle{Data-Driven Identification of Accidental Clicks on Mobile Ads with Applications to Advertiser Cost Discounting and Click-Through Rate Prediction}

\author{Gabriele Tolomei \and Mounia Lalmas \and Ayman Farahat \and Andrew Haines \thanks{All the authors contributed to this work while employed at Yahoo Research.}}
\institute{	Gabriele Tolomei \at University of Padua, Italy \\
			\email{gtolomei@emath.unipd.it}
\and
			Mounia Lalmas \at Spotify, London, UK
			\email{mounia@acm.org}
\and
			Ayman Farahat \at Amazon, Palo Alto, CA, USA \\
			\email{afarahat@amazon.com}
\and
			Andrew Haines \at Yahoo Research @ Oath, London, UK\\
			\email{haines@oath.com}
}

\maketitle

\begin{abstract}
In the \emph{cost per click} (CPC) pricing model, an advertiser pays an ad network \emph{only} when a user clicks on an ad; in turn, the ad network gives a share of that revenue to the publisher where the ad was impressed.
Still, advertisers may be unsatisfied with ad networks charging them for ``valueless'' clicks, or so-called \emph{accidental clicks}.
These happen when users click on an ad, are redirected to the advertiser website and bounce back without spending \emph{any} time on the \emph{ad landing page}.
Charging advertisers for such clicks is detrimental in the long term as the advertiser may decide to run their campaigns on other ad networks. In addition, machine-learned click models trained to predict which ad will bring the highest revenue may overestimate an ad click-through rate, and as a consequence negatively impacting revenue for both the ad network and the publisher.

In this work, we propose a data-driven method to detect accidental clicks from the perspective of the ad network. 
We collect observations of time spent by users on a large set of ad landing pages -- \ie, \emph{dwell time}.
We notice that the majority of per-ad distributions of dwell time fit to a \emph{mixture of distributions}, where each component may correspond to a particular type of clicks, the first one being accidental.
We then estimate dwell time \emph{thresholds} of accidental clicks from that component.

Using our method to identify accidental clicks, we then propose a technique that \emph{smoothly} discounts the advertiser's cost of accidental clicks at billing time.
Experiments conducted on a large dataset of ads served on Yahoo mobile apps confirm that our thresholds are stable over time, and revenue loss in the short term is marginal.
We also compare the performance of an existing machine-learned click model trained on all ad clicks with that of the same model trained only on non-accidental clicks. There, we observe an increase in both ad click-through rate (+3.9\%) and revenue (+0.2\%) on ads served by the Yahoo Gemini network
when using the latter. These two applications validate the need to consider accidental clicks for both billing advertisers and training ad click models.
\CRclass{Information systems~Traffic analysis \and Information systems~Online advertising \and Computing methodologies~Mixture modeling}
\end{abstract}

\section{Introduction}
\label{sec:intro}
The time a user spends on a web page, referred as \emph{dwell time}, varies considerably by user and type of content. 
Still, for a given web page, dwell time can be used as an effective \emph{proxy} of engagement with its content~\cite{BarbieriWWW2016,Goldman2014GSP,LalmasKDD2015,YiRecSys2014}. 
If the content is of no interest to the user or is presented poorly, the dwell time on the web page will often be short. 
By contrast, for a user engaged with the content the dwell time will be longer. 
A third case, which does not depend on the content or its presentation, corresponds to \emph{extremely} short dwell time. 
This paper focuses on these extremely short dwell times.

In online advertising, when a user clicks on an advertisement, or ad for short, she is redirected to the advertiser web page, \ie, the \emph{ad landing page}. The dwell time on the ad landing page is measured as the time between the ad click and the user returning to the publisher site where the ad was impressed. 
There are three \emph{types} of clicks:
\begin{itemize}
	\setlength\itemsep{0pt}
\item[--] \emph{accidental click}: The user clicks on the ad, likely by mistake, reaches the ad landing page, and immediately bounces back; the user spends no time on the landing page.
\item[--] \emph{short click}: The user intended to click on the ad but once on the landing page decides to bounce back;  the post-click experience is not  satisfactory, due to the low quality of the landing page or its low relevance.
\item[--] \emph{long click}: Once landing on it, the user engages with the ad landing page and spends time on the advertiser site.
\end{itemize}

\begin{figure}[t]
\begin{center}
\includegraphics[width=.9\columnwidth]{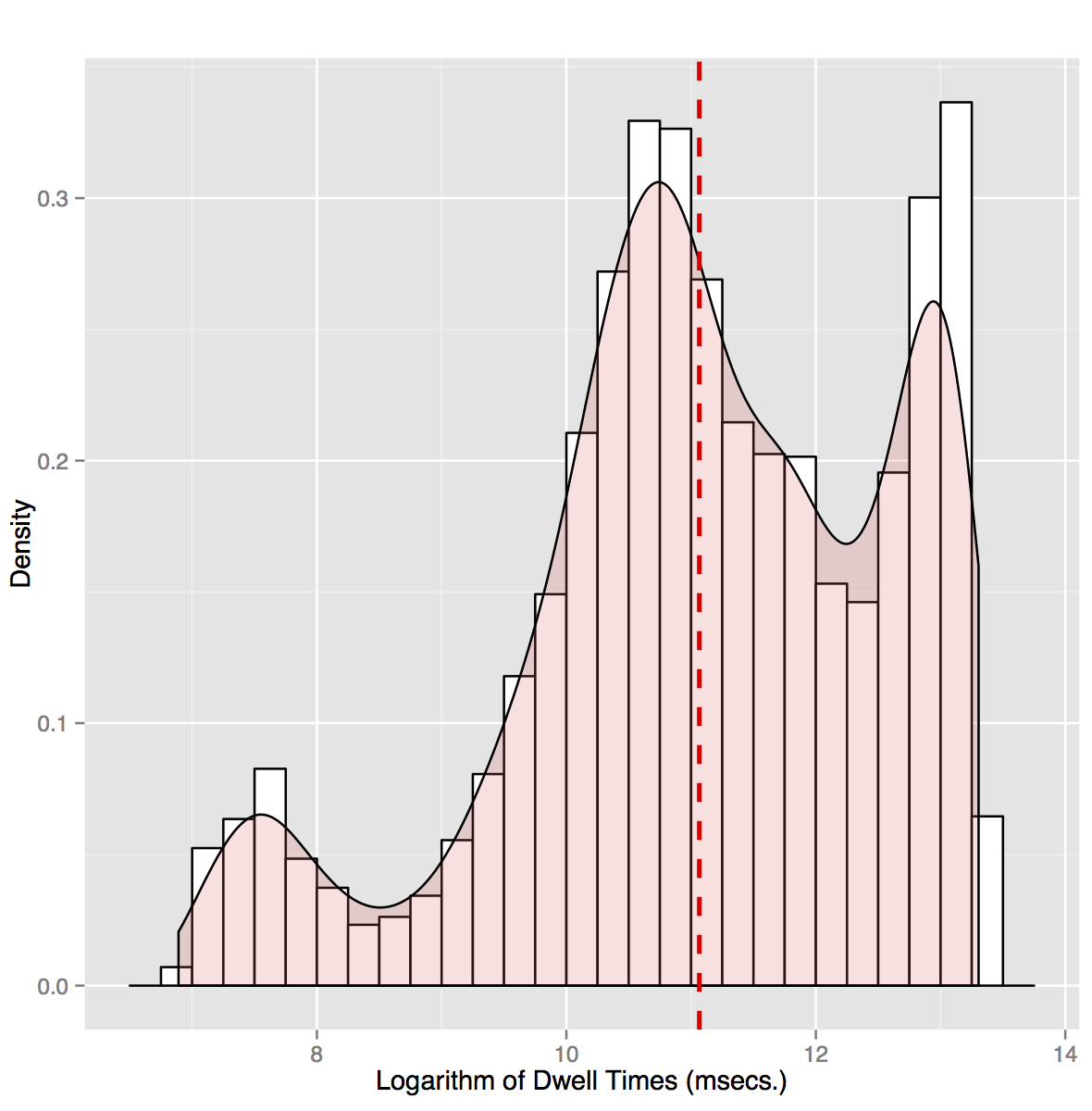}
\caption{\small{Empirical distribution of the logarithm of dwell time observed on an ad landing page.}}
\label{fig:dwelltime}
\end{center}
\end{figure}

We plot in Figure~\ref{fig:dwelltime} 
the distribution of the natural logarithm of dwell time values observed on an ad landing page.
The distribution is not unimodal; a small component can be identified with extremely low dwell time values, around $e^{7.5}\approx$ 1.8 seconds, as representative of accidental clicks, whereas the other two components capture the short and long clicks, respectively, demonstrating the existence of the above mentioned  three types of clicks.  
In this paper, we propose a data-driven approach for  estimating the dwell time \emph{thresholds} to identify whether a click is accidental or not. 

Properly accounting for accidental clicks is business-critical to online advertising. 
Consider the widely adopted \emph{cost per click} (CPC) pricing model, when an advertiser is charged by the ad network only when a user clicks on an ad.
Users may accidentally click on an ad, get redirected to the ad landing page and then bounce back without looking at it. 
This behaviour is even more severe on smartphones, as their limited screen size makes it more prone for users to click on ads by mistake~\cite{GoogleAdWords2016,StewartHCI2012}.
It is therefore not unusual for advertisers running CPC campaigns to complain when charged for such clicks, as these are ``valueless'' to them.
Ignoring these complaints can be detrimental in the long term, as it may affect the relationship between the ad network and the advertiser, who might at worst switch to other ad networks to run their campaigns.

A main challenge is for the ad network to accurately identify accidental clicks. Current solutions use hand-coded thresholds of dwell time on landing page to determine whether an ad click is accidental or not; for example, all visits to ad landing pages shorter than 5 seconds may be considered as accidental. With this approach, the threshold is fixed and set arbitrarily, and does not take into account empirically observed dwell time of ad clicks. Plus, while 5 seconds may be a reasonable threshold for ad clicks on a tablet, it may not be the right one for a smartphone. 
As a first and major contribution of this paper, we propose an unsupervised learning method that estimates thresholds of accidental clicks by fitting observed dwell time data to mixture models, which capture the three types of clicks defined above. 

Further, we deploy our method for identifying accidental clicks on two applications. 
The first one is concerned with a controlled approach to discount accidental clicks when charging advertisers. 
In principle, accidental clicks could be simply discarded once detected, and advertisers not charged for them. However, this strategy will negatively impact revenue for both the ad network and, consequently, the publisher.
Therefore, as a second contribution of this paper, we propose a \emph{smooth} discounting method based on the proportion of detected accidental clicks. 
The third contribution of this paper presents how accidental clicks identified by our approach are discarded from an existing machine-learned clicks model used to predict an ad \emph{click-through rate} (CTR for short). 
Our intuition is that by removing valueless clicks we can feed machine-learned models with ``cleaner'' training data, mitigating any possible overestimation of CTR. 

The rest of the paper is organised as follows. 
In Section~\ref{sec:dwellconv}, we motivate why we use dwell time as a proxy of an ad click value, and discuss why it is appropriate to adopt a data-driven solution for discovering accidental clicks.
Sections~\ref{sec:framework} and~\ref{sec:experiments} describe our data-driven approach to detect accidental clicks and experiments carried out with real-world data, respectively. 
Sections~\ref{sec:discount} and~\ref{sec:training} present two applications where our proposed approach to identify accidental click was deployed. In Section~\ref{sec:relwork}, we discuss related work and position our contributions. 
Finally, Section~\ref{sec:conclusions} concludes our work.
%
\section{Accidental clicks}
\label{sec:dwellconv}
We show first how we estimate the value of an ad click using dwell time, and then motivate using a data-driven approach to identify accident clicks.

\subsection{Using dwell time as the value of an ad click}
Advertisers may wish to be fully charged only for valuable clicks, \ie, those that lead to \emph{conversions},\footnote{\small{There is no consensus around what a \emph{conversion} is; it is up to the advertiser to specify it.}} whereas all  remaining clicks should be proportionally discounted.
Implementing this strategy requires an accurate estimate of the \emph{conversion rate}, \ie, the probability of conversion conditioned on a click.
However, conversion rate suffers from three major problems. 
First, it is hard to estimate since conversion data are often not available for a large number of advertisers.
Second, those data are not missing at random, as advertisers sharing their conversion data would be a biased sample. Finally, using conversion data to identify valuable clicks may lead to high false positive rates, since clicks not followed by conversions are not necessarily ``valueless''. In fact, those clicks represent profitable feedbacks for advertisers.  

It is therefore more appropriate to focus on identifying valuable clicks as clicks that are \emph{not} valueless, thus using so-called accidental clicks as complementary of conversions. 
We propose to use dwell time as our proxy measure of the value of an ad click. 
Using dwell time can help alleviate the aforementioned problems provided that dwell time is indeed a good proxy for conversion in our data. This was shown to be indeed the case in~\cite{Goldman2014GSP},  
and we further validate this in our context. 

\begin{table}[t]
	\centering 
	\small
	 \begin{tabular}{l*{4}{c}} \hline\hline
\multicolumn{1}{c}{Conversion}&\multicolumn{1}{c}{Mean}&\multicolumn{1}{c}{Std. Err.} &\multicolumn{1}{c}{Std. Dev.}  \\
\hline
{\tt yes}  &    5.729    &   .064  &    2.582 \\
{\tt no}  &    3.264 &    .006  &    2.569 \\
\hline
\end{tabular}
\caption{\small{Statistics on the natural logarithm of dwell time.\label{tab2}}}
\label{tab:summary}
\end{table}

Table~\ref{tab:summary} shows statistics on the natural logarithm of dwell time from two samples, one containing observations of dwell time leading to conversions ({\tt yes}) and the other made up of those that do not ({\tt no}) for 40 ads managed by a large ad network, Yahoo \emph{Gemini}\footnote{\small{\url{https://gemini.yahoo.com/}}}. 
We test against the null hypothesis that the mean of dwell time computed from the {\tt yes}-sample ($e^{5.729}\approx 307.7$ seconds) is the same as that calculated from the {\tt no}-sample ($e^{3.264}\approx 26.2$ seconds).
We run a two-tailed two-sample \emph{t-test} and are able to reject the null hypothesis stating that the difference of those two means is 0 (level of significance $\alpha = 0.01$, $p$-value $\ll 0.01$). 
Therefore,  the average dwell time of the two samples  are indeed statistically significantly different.

In addition, we test two regression models to verify if dwell time is a good predictor of conversion.
Let $Y_{i,j}$ be the binary random variable representing the conversion event occurring after the $j$-th click on the $i$-th ad.
We also denote by $X_{i,j}$ the logarithm of dwell time on the $i$-th ad landing page after the $j$-th click.
The simplest model we test is a linear regression of $Y_{i,j}$ on $X_{i,j}$, that is:
\begin{equation}
	\label{eq:linearreg}
Y_{i,j} = \beta_0 + \beta_1 X_{i,j} + \epsilon_{i,j}
\end{equation}
where $\epsilon_{i,j} \sim \mathcal{N}(0,\sigma^2_{i,j})$ is a zero-mean Gaussian error term.
A variant of this model simply applies the logit operator to the right-hand side of Equation~\ref{eq:linearreg}, namely:
\begin{equation}
	\label{eq:logitreg}
Y_{i,j} = \text{logit}(\beta_0 + \beta_1 X_{i,j}) + \epsilon_{i,j}
\end{equation}

\begin{table}[t]\centering 
	\small 
	\begin{tabular}{c|c|c}
                    & linear (Equation~\ref{eq:linearreg}) & logit (Equation~\ref{eq:logitreg})\\
\hhline{=|=|=}
$\beta_0$ (intercept)            &    -0.00384 &      -6.424  \\
$\beta_1$ (dwell time)            &     0.00413&       {\bf 0.301*}  \\
\hline
AIC & 14,399 & {\bf 14,297} \\
\hline
\end{tabular}
\caption{\small{Coefficients of linear and logit regression model.\label{tab1}}}
\label{tab:model}
\end{table}

Table~\ref{tab:model} shows the regression coefficients obtained with the two models above.
The best performing model is the logit, which is selected as the one with the smallest \emph{Akaike Information Criterion} (AIC)~\cite{Aikake1973AIC}.
We observe that the coefficient associated with the natural logarithm of dwell time ($\beta_1$) is both positive and significant, hence is a good predictor of conversion.
It is worth noting that a \emph{unit} increase in natural logarithm of dwell time (\ie, $\approx 2.72$ seconds, since 1 unit = $\text{ln}(x)$, where $x$ is dwell time in seconds) will increase the odds of conversion by 30\%.
This provides further justification in using dwell time as proxy of conversion, and therefore of ad click value.

\subsection{A data-driven approach to identifying accidental clicks}

Using dwell time as proxy of the ``value'' of an ad click, this paper puts forward a data-driven approach -- based on actual dwell time observations -- to identify whether an ad click is accidental or not.
On the other hand, using fixed thresholds on dwell time to identify accidental clicks -- say 1 second -- is instead a common practice. The reason being is that this is straightforward, as no data processing nor computation is required, and can be easily integrated with any existing production systems. This simple approach however 
prevents capturing subtleties arising from hidden latent factors, such as the device (desktop \vs\ mobile), the application (\eg, mail \vs\ news stream) or the network type (\eg, 3G/4G \vs\ Wi-Fi).
A data-driven approach can account for all these factors, thus providing more reliable \emph{estimates} of dwell time thresholds of accidental clicks from observed data.
Moreover, the analysis of observed dwell times may  characterise other phenomena; 
\eg, not only the lowest values (accidental clicks) but also  large and extremely large ones (short and long clicks, respectively). We leave this for future work, as we focus on accidental clicks in this paper.
%
\section{Discover accidental clicks}
\label{sec:framework}

Our data-driven approach for detecting accidental clicks on ads consists of two steps: \emph{data modelling} and \emph{dwell time thresholding}.
The former fits observations of dwell time of a large set of ad landing pages to a probabilistic model, whereas the latter estimates threshold of dwell time to identify accidental clicks. 
We employ an unsupervised learning approach, as no supervised learning approaches to classify ad clicks as accidental can be designed as this would require building a ground truth, which is not achievable.

\subsection{Data modelling}

We assume that observations of dwell time of ad clicks are generated by an underlying probabilistic model.
Ideally, such a model has to simultaneously represent the three types of clicks shown in Figure~\ref{fig:dwelltime},  \emph{accidental}, \emph{short}, and \emph{long}.

Generally speaking, a \emph{mixture of distributions} is a probabilistic model that captures the presence of ``sub-po\-pu\-la\-tions'' within an overall population~\cite{Lindsay1995,McLachlan2000}. As such, it is a good candidate to describe our observations of dwell time.
More formally, a continuous random variable $X$ (\eg, dwell time) is distributed according to a mixture of $K$ (discrete) component distributions if its probability density function (pdf) $f_X$ is a convex combination of $K$ pdfs $f_1, \ldots, f_K$:
\begin{equation}
\label{eq:mixture}
f_X(x;\bm{\Theta}) = w_1 f_1(x; \bm{\theta_1}) + \ldots + w_K f_K(x; \bm{\theta_K}) = \sum_{i=1}^K w_i f_i(x; \bm{\theta_i})
\end{equation}
where: 
\begin{itemize}
	\setlength\itemsep{0em}
\item[--] each $f_i$ belongs to the same (parametric) family of distributions (\eg, Normal,\footnote{\small{We refer to \emph{Normal} and \emph{Gaussian} distribution, interchangeably.}} Log-Normal, Gamma, Weibull, \etc);
\item[--] $w_i$ is the \emph{mixture weight} (or prior probability) associated with the $i$-th component;
\item[--] $w_i \geq 0$ and $\bm{w} = (w_1, \ldots, w_K)^T$ is the $K$-dimensional vector of mixture weights, so that $\sum_{i=1}^K w_i = 1$;
\item[--] $\bm{\theta_i}$ is the vector of \emph{parameters} associated with the $i$-th component, \eg, if $f_i$ is the pdf of a Normal distribution $\mathcal{N} (\mu_i,\sigma_i^2)$ then $\bm{\theta_i} = (\mu_i, \sigma_i^2)$;
\item[--] $\bm{\Theta} = (\bm{w}, \bm{\theta_1}, \ldots, \bm{\theta_K})$ is the \emph{overall} vector of parameters of the mixture model;
\item[--] there exists a \emph{latent random variable} denoted by $I$ governing which component each observation of $X$ is drawn from. This random variable is distributed according to a categorical distribution whose parameter is the vector of mixture weights $\bm{w}$, so that:
\begin{enumerate}
	\setlength\itemsep{.0em}
\item $I \sim \text{Categorical}(\bm{w}) \longmapsto$ pick the component distribution $f_i$ with probability $w_i$;
\item $X=x'~|~I=i\longmapsto $ generate a value for $X$ from the component distribution $f_i$.
\end{enumerate}
\end{itemize}
We describe next how to model dwell time on ad landing pages using a mixture of distributions.

\subsubsection{Mixture of distributions for dwell time}

Representing the three types of clicks can be done by having the observed dwell times on ad landing pages generated by a mixture of $K=3$ distributions. Let $M$ be the number of ads.
For each ad $j \in \{1, \ldots, M\}$ we consider a sample $\mathcal{D}_j$ of $n_j$ i.i.d.~\emph{positive} random variables $X_{j,1}, \ldots, X_{j,n_j}$, with each $X_{j,k}$ representing an observation of the dwell time associated with the $k$-th click on the ad $j$: $\mathcal{D}_j = \{X_{j,1}=x_{j,1},\ldots, X_{j,n_j}=x_{j,n_j}\}$.
Each $X_{j,k}$ is drawn from a mixture of up to $K=3$ components, so that the pdf of $X_{j,k}$ is:
\begin{equation}
\label{eq:threecomp}
f_{X_{j}}(x;\bm{\Theta}) = \sum_{i=1}^3 w_i f_i(x; \bm{\theta_i})
\end{equation}
In addition, each $f_i$ is the pdf of the same parametric distribution, although we only consider probability distributions with positive domain (\eg, Log-Normal, Gamma, Weibull, \etc) as dwell time cannot be negative.

Next, we discuss how the parameters $\bm{\Theta}$ of the mixture model can be estimated from the observed data.

\subsubsection{Parameter estimation}
For each ad, we estimate the overall vector of parameters $\bm{\Theta} = (\bm{w}, \bm{\theta_1}, \bm{\theta_2}, \bm{\theta_3})$ from the observed dwell times using \emph{maximum likelihood estimation} (MLE).
For each ad $j$ we know that the pdf of each of its observations is a mixture of three distributions, defined as in Equation~\ref{eq:threecomp}.
Since all observations in $\mathcal{D}_j$ are independent and identically distributed, we compute their joint probability density $f_{X_j}$ as:
\begin{equation}
\label{eq:jpd}
f_{X_j}(\mathcal{D}_j;\bm{\Theta}) = f_{X_j}(x_{j,1},\ldots,x_{j,n_j};\bm{\Theta}) = \prod_{k=1}^{n_j} f_{X_j}(x_{j,k};\bm{\Theta})
\end{equation}
From the joint probability density we derive the \emph{likelihood function} $L(\bm{\Theta};\mathcal{D}_j)$, as $L(\bm{\Theta};\mathcal{D}_j) = f_{X_j}(\mathcal{D}_j;\bm{\Theta})$.
Although the two functions are the same, the likelihood function emphasises 
that the dataset is fixed and the parameters $\bm{\Theta}$ are variable.
The aim of MLE is thus to find a value of ${\boldsymbol \Theta}$ -- \ie, an estimate $\hat{{\boldsymbol \Theta}}_{\text{MLE}}$ --  maximising the likelihood function:\footnote{\small{In practice, we often seek for $\hat{{\boldsymbol \Theta}} _{\text{MLE}}$ so as to maximise the \emph{log-likelihood function} $\text{ln}(L(\bm{\Theta};\mathcal{D}_j))$, since this is equivalent (the natural logarithm is monotonically increasing) but simpler because products change into summations.}}
\begin{eqnarray}
\label{eq:mle}
\hat{{\boldsymbol \Theta}}_{\text{MLE}} &=& \text{argmax}_{\boldsymbol \Theta}\Big[L(\bm{\Theta};\mathcal{D}_j)\Big ] \nonumber \\
&=&\text{argmax}_{\boldsymbol \Theta}\Big[\prod_{k=1}^{n_j} f_{X_j}(x_{j,k};\bm{\Theta})\Big ] \nonumber \\ 
&=&\text{argmax}_{\boldsymbol \Theta}\Big[ \prod_{k=1}^{n_j} \sum_{i=1}^3 w_i f_i(x_{j,k}; \bm{\theta_i}) \Big]
\end{eqnarray}
Different likelihood functions to be maximised can be obtained depending on which pdf we fill in Equation~\ref{eq:mle} with.
However, if the resulting $L(\bm{\Theta};\mathcal{D}_j)$ is differentiable in $\boldsymbol \Theta$, we can find $\hat{{\boldsymbol \Theta}}_{\text{MLE}}$ as a solution of the system of equations:
\begin{equation}
	\label{eq:gradient}
\nabla L(\bm{\Theta};\mathcal{D}_j) = \frac{\partial{L(\bm{\Theta};\mathcal{D}_j)}}{\partial{\theta}} = 0, \theta \in \bm{\Theta}
\end{equation}
where $\nabla$ is the \emph{gradient} of the likelihood function, \ie, the vector of partial derivatives of the likelihood function with respect to each parameter in $\bm{\Theta}$.

Unfortunately, a maximum likelihood estimation of the parameters ${\boldsymbol \Theta}$ is not straightforward, since often there are no closed-form solutions to Equation~\ref{eq:gradient} available; as such, we cannot solve directly $\nabla L(\bm{\Theta};\mathcal{D}_j) = 0$~\cite{SchlattmannMAFMM2009}.
A typical solution is to use the \emph{expectation-max\-imi\-za\-tion} (EM) algorithm~\cite{BishopPRML2006}.
EM is an \emph{iterative}, numerical approximation procedure that starts with an initial random guess for the values of the parameters and converges to a local maximum (or to a saddle point) of the observed-data likelihood.
Although EM does not guarantee convergence to a global maximum, in practice there are a variety of heuristic approaches for escaping a local maximum: multiple restarts, clever initialization, and modifications to the EM algorithm itself~\cite{ElkanMM2010}.
Finally, for each ad we compute the set of model parameters $\hat{{\boldsymbol \Theta}}_{\text{MLE}}$ which maximise the likelihood function.

\subsubsection{Model selection}

To choose the ``best'' model for each ad, we cannot just select the one with parameters $\hat{{\boldsymbol \Theta}}_{\text{MLE}}$, \ie, the one that best fits to the observed data.
In general the more complex (flexible) is the model the better will be its goodness-of-fit to the observed data; in other words, the higher will be its likelihood function computed with respect to the observed data.
At the same time, the more complex (flexible) is the model the less it generalises to unseen data; in other words, the higher is the chance of the model to \emph{overfit} the observed data~\cite{JamesISL2014}. 
Therefore, if we choose the model having the highest likelihood we always end up selecting the one having the maximum degree of freedom, \ie, the maximum number of components $K=3$.

Therefore, to avoid overfitting and find a trade-off between complexity and interpretability\footnote{\small{This is also often referred to as the \emph{bias-variance} trade-off.}} we use tools such as the \emph{Akaike Information Criterion} (AIC)~\cite{Aikake1973AIC} or \emph{Bayesian Information Criterion} (BIC)~\cite{Schwarz1978BIC}.
The former is computed as $\text{AIC} = - 2\ln(L) + 2K$, whereas the latter as $\text{BIC} = - 2\ln(L) + K\ln(n)$, where $K$ is the number of components of the model, $L$ is the likelihood function as maximised by the parameters of the model estimated from the observed data, and $n$ is the dataset size.
Both criteria try to penalize models that are unnecessarily too complex, and finally select the one with the \emph{smallest} AIC or BIC.

So far, we have estimated and selected the mixture model that best describes the observed dwell times on each ad landing page.
Next, we present how we can use this model to compute dwell time thresholds to identify accidental clicks. 

\subsection{Dwell time threshold of accidental clicks}
For each ad, we fit the observed dwell times on its landing page to a mixture of distributions using MLE and one of the model selection criteria described above.
We then focus on the subset of ads exhibiting exactly \emph{all} three components, namely ads with dwell times fitting a mixture of three distributions.
Intuitively, these are the ads showing all the three categories of clicks we have conjectured the existence of, namely, \emph{accidental}, \emph{short} and \emph{long}. 

Given an ad and the set of parameters of all its components, we compute statistics such as the expected value or the median of every component.
As we are interested in detecting accidental clicks we only focus on the \emph{first} component of each ad. Using the second and third component to study short and long clicks, respectively, is something we leave for future work.

For example, if we fit the data to a mixture of three Log-Normal distributions we can represent the first component by a random variable distributed as a Log-Normal with parameters $\bm{\theta_1} = (\mu_1, \sigma_1^2)$.
In general, for any random variable $Z\sim \ln\mathcal{N}(\mu, \sigma^2)$, or equivalently $\ln(Z) \sim \mathcal{N}(\mu, \sigma^2)$, we can derive the following:
\begin{itemize}
	\setlength\itemsep{0em}
\item[--] $E[Z] = e^{\mu + \sigma^2/2}$ (where $E$ denotes the expected value);
\item[--] $\text{Median}(Z) = e^{\mu}$.
\end{itemize}
We therefore estimate a per-ad threshold of dwell time for detecting accidental clicks using either the expected value or the median of the first component -- the latter being more robust to the presence of outliers -- by letting $\mu = \mu_1$ and $\sigma^2 = \sigma_1^2$ in the equations above.
Finally, to obtain an overall estimate of dwell time threshold of accidental clicks across all the ads, we compute the mean or the median of the individual per-ad estimates.

In this section, we described a two-stage approach for computing thresholds of dwell time from observed data, to detect accidental clicks on ads.
Next, we present experimental results when these thresholds are deployed within a large ad network. 
%
\section{Experiments}
\label{sec:experiments}

We conduct two experiments on multiple datasets of ads served by a large ad network, codenamed \emph{Gemini}, on several Yahoo mobile apps.
We focus on mobile apps as these are where accidental clicks are more likely to happen~\cite{GoogleAdWords2016,StewartHCI2012}. We only  consider ads with at least 100 clicks to increase the confidence of the estimates of our thresholds.

In the first experiment, we choose one \emph{pivoting} mobile app to estimate the dwell time threshold of accidental clicks, which is then used to identify accidental clicks on other (two in our case) mobile apps. 
To protect sensitive information, we refer to the former as {\sf App 1} and to the other two as {\sf App 2} and {\sf App 3}, respectively.
The experiment is performed on two one-month datasets, which we refer to $\mathcal{D}_{1}$ and $\mathcal{D}_{2}$, respectively.
Each consists of a random sample of around 10,000 ads and 6.5M clicks, unevenly distributed on the three mobile apps.
In the second experiment, instead of having one pivoting mobile app used to estimate a \emph{single} dwell time threshold of accidental clicks, we generate a threshold for each mobile app. The dataset used is a random sample from three weeks worth of data containing 120,000 ads and  70M clicks, hereinafter called $\mathcal{D}_{3}$. 

With the first approach -- using a pivoting app -- the aim is provide a single estimate of dwell time threshold of accidental clicks, using the app with the highest volume of traffic. The second approach provides one threshold per app, which is more flexible and accounts for the effect of different user experience, population, and operating systems (\ie, Android \vs\ iOS). 

\subsection{Data preprocessing}
\label{sec:discovery:data}
In both experiments, we remove outliers by discarding clicks with dwell time greater than 600 seconds.\footnote{\small{A threshold already used in previous work~\cite{LalmasKDD2015}.}}
We also apply a logarithmic transformation to all the observations. 
This allows us to fit the log-transformed data to a mixture of Gaussian distributions, as this is the same as fitting the original data to a mixture of Log-Normals.

The logarithmic transformation emphasises differences between small values of dwell time, whereas it smooths the same differences when they happen between larger values of dwell time.
In this way, we are able to capture relative differences instead of absolute ones. 
To identify accidental clicks a difference of 1 second between two small values of dwell time, such as 2 and 3 seconds, is more important than the same difference between, say, 101 and 100 seconds.

Our approach is not tailored to any specific family of parametric distributions, and we select a mixture of Log-Normal distributions purely because this gives us the best (smallest) AIC on average over all the ads. 

\subsection{Single threshold from a pivoting app}
\label{subsec:singleth}
\begin{figure*}[t]
        \centering
        \begin{subfigure}[b]{0.32\textwidth}
                \includegraphics[width=\textwidth]{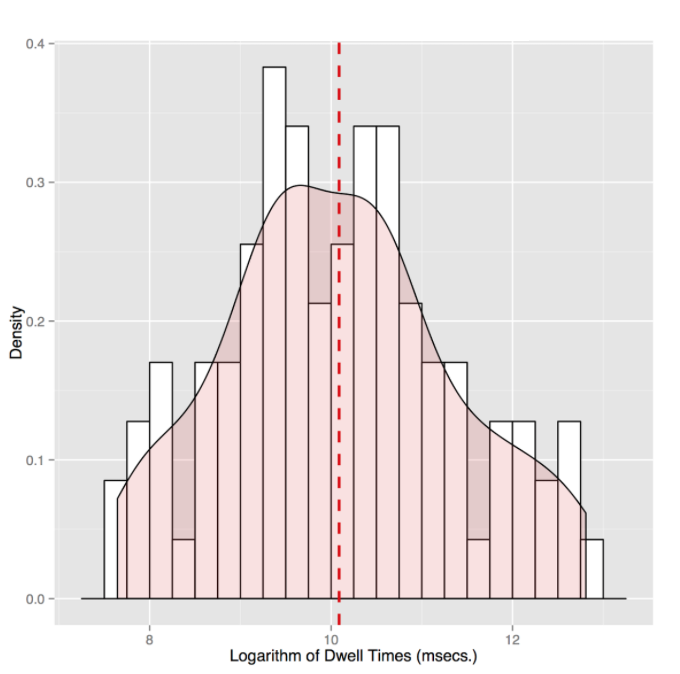}
                \caption{\small{One component}}
                \label{fig:1-comp}
        \end{subfigure}%
        ~ 
        \begin{subfigure}[b]{0.32\textwidth}
                \includegraphics[width=\textwidth]{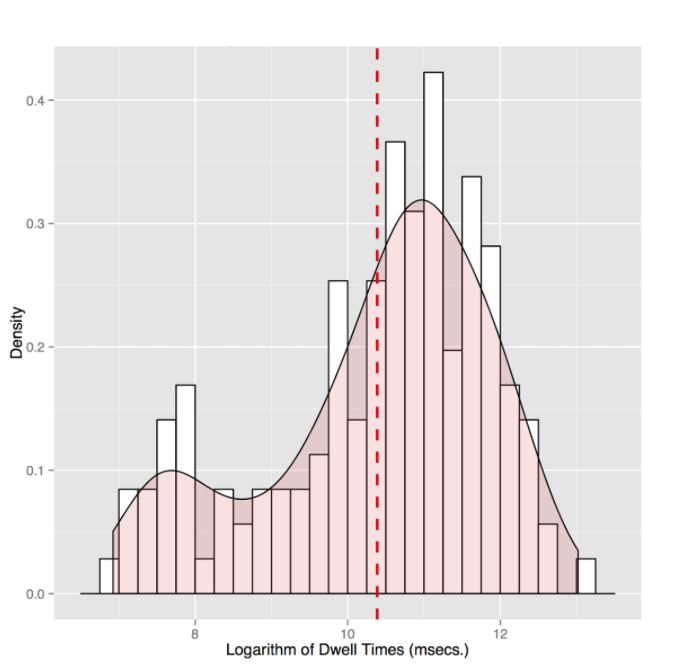}
                \caption{\small{Two components}}
                \label{fig:2-comp}
        \end{subfigure}
        ~ 
        \begin{subfigure}[b]{0.32\textwidth}
                \includegraphics[width=\textwidth]{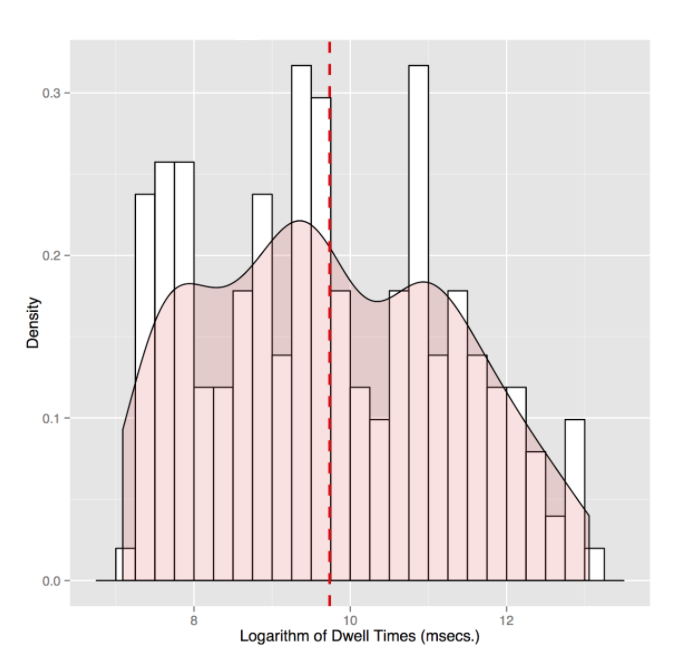}
                \caption{\small{Three components}}
                \label{fig:3-comp}
        \end{subfigure}
        \caption{\small{Examples of ads clicked on the pivoting {\sf App 1} which fit to one, two, and three components.}}\label{fig:comp}
\end{figure*}

In this first setting, we consider (log-trans\-formed) observations of dwell time from the ads clicked on our pivoting mobile app -- {\sf App 1} -- collected both from $\mathcal{D}_{1}$ and $\mathcal{D}_{2}$.
Then, we fit each set of ad dwell time observations to a mixture of Gaussian distributions. 
Figure~\ref{fig:comp} presents three examples of ads, each one fitted to mixture of one, two, and three Gaussian distributions.
Except for the ad shown in Figure~\ref{fig:1-comp}, the other two exhibit a first component centered around very small value of dwell time (around $e^{7.7} \approx 2.2$ seconds), which likely represents accidental clicks.

Table~\ref{tab:datafit} shows how ads in the two datasets fit to models having one, two, and three components, respectively. For both, the vast majority of ads (82.5\% and 65.4\%, respectively) fit to exactly three components.
These ads are the ones we focus on to detect the dwell time threshold of accidental clicks, since their corresponding landing pages contain all the three categories of clicks described in Section~\ref{sec:intro}.
Since our aim is to ``isolate'' accidental clicks happening on our pivoting app, we concentrate on the first component.
According to our conjecture, this would capture users clicking on an ad by mistake, or simply returning to the publisher site without actually landing on the advertiser's page.

\begin{table}[t]
\centering
\small
\begin{tabular}{cc|c|cl}
\cline{2-4}
& \multicolumn{1}{ |c }{{\bf1 comp}} & \multicolumn{1}{ |c }{{\bf2 comps}} & \multicolumn{1}{ |c| }{{\bf3 comps}}\\
\cline{1-4}
\multicolumn{1}{ |c }{$\mathcal{D}_{1}$} & \multicolumn{1}{ |c }{1.0\%} & \multicolumn{1}{ |c }{16.5\%} & \multicolumn{1}{ |c| }{82.5\%}     \\ \cline{1-4}
\multicolumn{1}{ |c }{$\mathcal{D}_{2}$} & \multicolumn{1}{ |c }{2.9\%} & \multicolumn{1}{ |c }{31.7\%} & \multicolumn{1}{ |c| }{65.4\%}     \\ \cline{1-4}
\end{tabular}
\caption{\small{Percentage of ads clicked on the pivoting {\sf App 1} which fit to one, two, and three components.}}
\label{tab:datafit}
\end{table}

For each ad we compute an estimate of the dwell time threshold using the median of its first fitted component, as this is more robust to the variance.
In fact, we observe that the variance increases going from the first to the second and finally to the third component.
Intuitively, this reflects the variability of dwell time on each click category: dwell times of accidental clicks are expected to differ less between each other than what would be the case with short and long clicks.

To obtain an overall estimate of the threshold $t_{\text{acc}}$ (an estimate derived from all the per-ad estimates), we propose two strategies: \emph{(i)} the mean of all the per-ad medians; \emph{(ii)} the median of all the per-ad medians.
The first one results in a generally higher threshold, which implies considering ``accidental'' a larger number of clicks.
The second estimate is more ``conservative'' and usually generates a smaller value of the threshold.

\begin{figure*}[t]
        \centering
        \begin{subfigure}[b]{0.45\textwidth}
                \includegraphics[width=\textwidth]{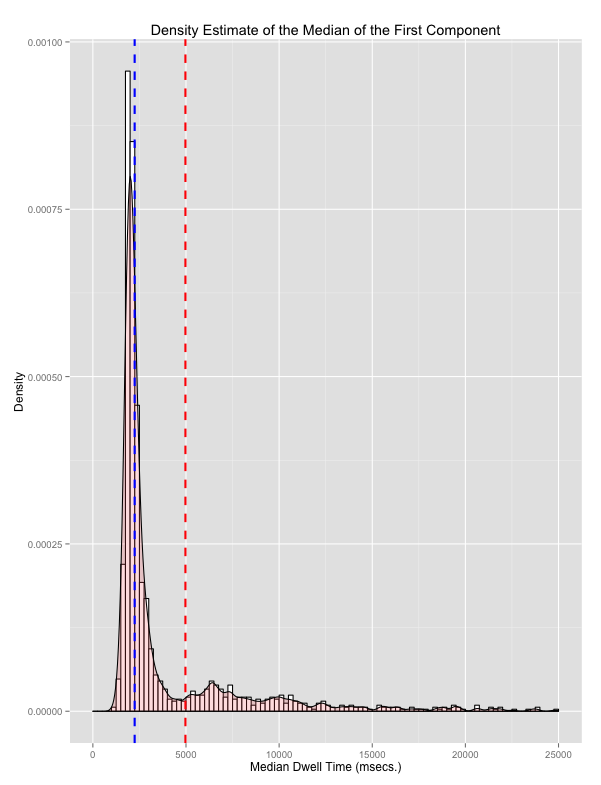}
                \caption{\small{$\mathcal{D}_{1}$}}
                \label{fig:density1}
        \end{subfigure}%
        ~ 
        \begin{subfigure}[b]{0.45\textwidth}
                \includegraphics[width=\textwidth]{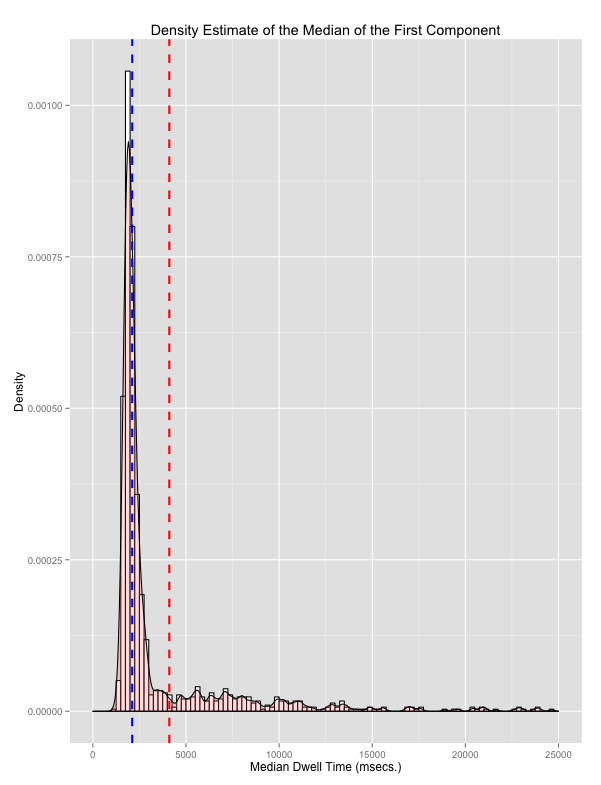}
                \caption{\small{$\mathcal{D}_{2}$}}
                \label{fig:density2}
        \end{subfigure}
        \caption{\small{Distributions of per-ad medians computed from the pivoting {\sf App 1} ({\color{blue}{median}} vs. {\color{red}{mean}}).}}\label{fig:median}
\end{figure*}

In Figure~\ref{fig:median}, we plot the distribution of per-ad medians of the first component computed from the pivoting {\sf App 1} on $\mathcal{D}_{1}$ and $\mathcal{D}_{2}$. The median of all those medians (the {\color{blue}{blue}} dashed line) seems more suitable than the mean (the {\color{red}{red}} dashed line), as it perfectly aligns with the ``peak'' we are interested in, which sits around $2.1 \div 2.2$ seconds. For business confidentiality, we do not disclose the percentage of accidental clicks for each of the considered apps, but we can report that this percentage is stable over the two datasets, once the thresholding strategy is fixed. Anecdotally, this percentage was shown to change using a dataset from a different time period, as the result of a change of the user interface on a specific app.

\subsection{Multiple per-app thresholds}
\label{subsec:multipleth}

In the previous setting, we reported results obtained when using a pivoting mobile app to compute a single threshold of dwell time, which in turn can be used to identify accidental clicks for other apps. 
In this section,  we discuss another approach, which was rolled out in production.
Instead of computing a single threshold from one pivoting app, we generate a dwell time threshold of accidental clicks for each app, individually. 
By doing so, we are also accounting for the effect of different user experiences or user populations on the different  mobile apps and operating systems (\ie, Android \vs\ iOS). 
We use a default threshold value for apps with not enough observations of dwell time. 

\begin{figure*}[t]
        \centering
        \begin{subfigure}[b]{0.32\textwidth}
                \includegraphics[width=\textwidth]{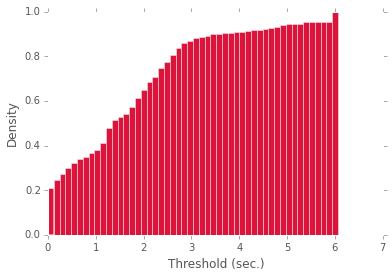}
                \caption{\small{First week}}
                \label{fig:1-week}
        \end{subfigure}%
        ~ 
        \begin{subfigure}[b]{0.32\textwidth}
                \includegraphics[width=\textwidth]{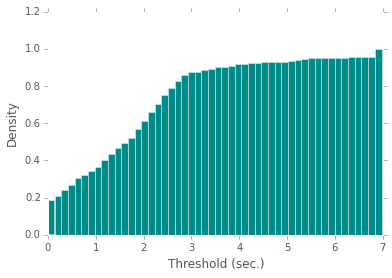}
                \caption{\small{Second week}}
                \label{fig:2-week}
        \end{subfigure}
        ~ 
        \begin{subfigure}[b]{0.32\textwidth}
                \includegraphics[width=\textwidth]{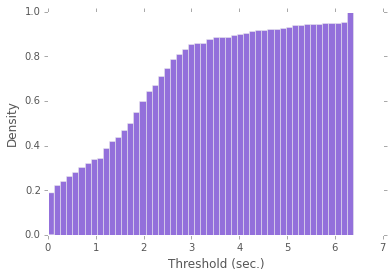}
                \caption{\small{Third week}}
                \label{fig:3-week}
        \end{subfigure}
        \caption{\small{Empirical CDF of thresholds of accidental ad clicks on mobile apps computed on $\mathcal{D}_{3}$.}}\label{fig:ecdf_thresholds}
\end{figure*}

In Figure~\ref{fig:ecdf_thresholds}, we plot the empirical cumulative distribution function (eCDF) of the thresholds computed on each week of the dataset ($\mathcal{D}_{3}$).
We observe that the distribution of thresholds remains stable over time. 
The median of all the per-app thresholds identified with this approach results in a value of 2.1 seconds, which aligns with the threshold generated using the pivoting strategy.

\begin{figure*}[htb!]
        \centering
        \begin{subfigure}[b]{0.45\textwidth}
                \includegraphics[width=\textwidth]{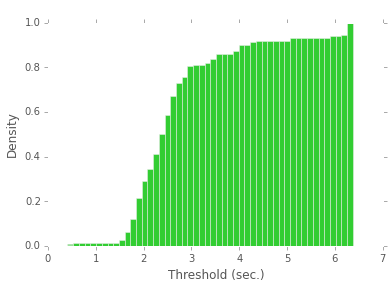}
                \caption{\small{Mobile Android Apps}}
                \label{fig:android}
        \end{subfigure}%
        ~ 
        \begin{subfigure}[b]{0.45\textwidth}
                \includegraphics[width=\textwidth]{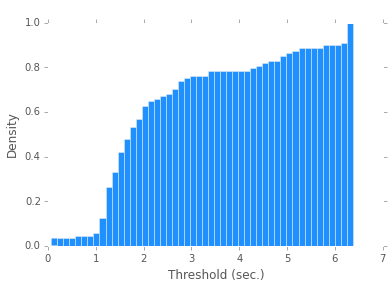}
                \caption{\small{Mobile iOS Apps}}
                \label{fig:ios}
        \end{subfigure}
        \caption{\small{Empirical CDF of thresholds of accidental ad clicks computed on $\mathcal{D}_3$ across different platforms.}}\label{fig:ecdf_platforms}
\end{figure*}

In Figure~\ref{fig:ecdf_platforms}, we separate the eCDFs of the thresholds obtained from apps on Android and iOS. 
We see that there are differences, as the median values are now $\approx 2.44$ and $\approx 1.80$ seconds, respectively.
In general, thresholds of accidental ad clicks on Android apps are more right-skewed than those computed for iOS apps, thereby suggesting that thresholds on Android are somewhat less dependent from the app which they are computed on. 

We presented two strategies to calculate dwell time thresholds of accidental clicks. Both strategies rely on the same data-driven approach, \ie, estimating the parameters of a mixture of three dwell time distributions and computing aggregated statistics (\eg, the median) on the first component. The two strategies differ in the (historical) dwell time data used to fit the mixtures. One uses observations of dwell time \emph{only} from a single app, and hence provides the threshold for apps  with few historical dwell time observations, who may be less popular than the pivoting app or have just entered the market. The other provides a per-app threshold, which can be used when there are multiple apps with a sufficiently large number of dwell time observations. In addition, this strategy considers the impact of different user experience on those apps.

In the next two sections, we present two use cases where our proposed data-driven approach for identifying  accidental clicks was deployed.
%
\section{Use case I: Discounting accidental clicks when billing advertisers}
\label{sec:discount}

With a mechanism for detecting accidental clicks, an ad network may simply discard all accidental clicks when billing the advertiser. 
This can however severely impact revenue for both the ad network and the publisher, at least in the short term. For example, in our dataset $\mathcal{D}_{3}$, we saw that 
the top-3 most revenue-losing apps account for $\approx 72.4\%$ of the overall revenue loss for all the apps under consideration.\footnote{\small{The actual revenue loss is not shown due to business confidentiality.}}
It is therefore important to control how much revenue loss is acceptable, hence looking for a trade-off between accounting for accidental clicks (satisfying the advertisers) and containing revenue loss (satisfying the ad network and publishers).
We present a smooth method to \emph{discount} the price of accidental clicks, instead of discarding them, so that advertisers are not fully charged for those clicks.

\subsection{Smooth discounting strategy}
\label{subsec:smoothdiscount}

One of the main attractions of ad networks is scale; advertisers have access to a large number of impressions and reach a wide audience with a single buy. 
However, not all the apps in the network perform equally. 
The advertiser is then faced with the problem of either selecting which apps to bid on or adjusting the bids by app. 
Both cases create extra friction for the advertiser. 
The algorithmic discounting we present below addresses this problem by adjusting the cost per click on each app of the network, such that the return on investment (ROI) for the advertiser is the same across all the apps.

We assume the existence of a \emph{pivoting} app from which we estimate the dwell time threshold of accidental clicks $t_{\text{median}}$, such as discussed in Section 4.2. 
Let $C_{i,\textsf{pivot}}$ be the set of clicks observed on ad $i$, which have been impressed on the pivoting app on a fixed time window.
Moreover, let $clicks_{i,\textsf{pivot}} = |C_{i,\textsf{pivot}}|$ be the total number of observed clicks on $i$.
Therefore, $C_{i,\textsf{pivot}}^{\text{non-acc}} = \{c \in C_{i,\textsf{pivot}}~|~\tau(c) > t_{\text{median}}\}$ -- where $\tau(c)$ is the dwell time of $c$ -- is the set of \emph{non}-accidental clicks on $i$ identified using $t_{\text{median}}$ on the same time window, and $clicks_{i,\textsf{pivot}}^{\text{non-acc}} = |C_{i,\textsf{pivot}}^{\text{non-acc}}|$ is the total number of \emph{non}-accidental clicks on $i$.
Similarly, for any other app we define $clicks_{i,\textsf{app}}$ and $clicks_{i,\textsf{app}}^{\text{non-acc}}$.

An advertiser may associate a \emph{value} to each click on ad $i$, referred to as $vfc_i$. This corresponds to 
the amount of money the advertiser would like to earn from a click on $i$, \emph{independently} of the source (app, in this case) where such click occurs.
Under a CPC cost model, there is a maximum amount of money the advertiser is willing to pay for having ad $i$ impressed and clicked, denoted by $cpc_i$. 
We define the advertiser ROI for ad $i$ on the generic {\sf app} as:
\begin{equation}
	\label{eq:roi}
	\frac{vfc_{i} \times clicks_{i,\textsf{app}}^{\text{non-acc}}}{cpc_{i} \times clicks_{i,\textsf{app}}}
\end{equation}
where the numerator is the total value earned by the advertiser considering only valid -- non-accidental -- clicks on {\sf app}, and the denominator is the total cost the advertiser would pay for \emph{all} the clicks on ad $i$ occurred on {\sf app}.
If we knew what is the \emph{true} non-accidental click rate of the app ($NACR_{\textsf{app}}$), we could rewrite Equation~\ref{eq:roi} as:
\begin{equation}
\frac{vfc_{i} \times clicks_{i,\textsf{app}}^{\text{non-acc}}}{cpc_{i} \times clicks_{i,\textsf{app}}} \approx \frac{vfc_{i} \times NACR_{\textsf{app}}}{cpc_{i}}
\end{equation}
Indeed, $\frac{clicks_{i,\textsf{app}}^{\text{non-acc}}}{clicks_{i,\textsf{app}}}$ is the MLE estimate of the true $NACR_{\textsf{app}}$.
We require the app we chose as pivot not only to be the one from which we can accurately estimate $t_{\text{median}}$ but also the best performing app with the highest $NACR$, \ie, the highest proportion of valid ad clicks overall. 
This is because we want the pivoting app to be the ``benchmark'' against which we compare all the other apps of the network.

The advertiser ROI calculated on any app of the network should be ideally equal to that of the pivoting app:
\begin{equation}
	\label{eq:roi2}
	\frac{vfc_{i} \times NACR_{\textsf{app}}}{cpc_{i}} = \frac{vfc_{i} \times NACR_{\textsf{pivot}}}{cpc_{i}}
\end{equation}
Moreover, since the value that the advertiser would get from a click on ad $i$ ($vfc_i$) is independent of the source, we can rewrite Equation~\ref{eq:roi2} as:
\begin{equation}
	\label{eq:roi3}
	\frac{NACR_{\textsf{app}}}{cpc_{i}} = \frac{NACR_{\textsf{pivot}}}{cpc_{i}}
\end{equation}
We may observe that $\frac{NACR_{\textsf{app}}}{NACR_{\textsf{pivot}}} \leq 1$, since $NACR_{\textsf{pivot}}$ is the highest among all the apps by design.
For Equation~\ref{eq:roi3} to be satisfied, we define $cpc'_{i,\textsf{app}} \leq cpc_{i}$ as the \emph{adjusted} cost of each accidental click on ad $i$, specific to {\sf app}:
\begin{equation}
	\label{eq:discountedcost}
	cpc'_{i,\textsf{app}} = \frac{NACR_{\textsf{app}}\times cpc_{i}}{NACR_{\textsf{pivot}}}
\end{equation}
The intuition is that the cost of an accidental click on $i$ on a generic app ($cpc'_{i,\textsf{app}}$) should be obtained by discounting the price for a valid click ($cpc_{i}$) proportionally to its relative value with respect to the best performing pivoting app of the network $\Big(\frac{NACR_{\textsf{app}}}{NACR_{\textsf{pivot}}}\Big)$, which is exactly our \emph{discount factor}.

This strategy does not discount accidental ad clicks on the pivoting app itself.
This is because the pivoting app is chosen as the one with the smallest accidental click rate, thus likely with little need to apply a discount factor.
Nonetheless, we may also decide to account for accidental clicks on the pivoting app, especially if its click value performance deteriorates.
Various strategies may be deployed.
For example, we can monitor the number of accidental clicks on the pivoting app and if it eventually exceeds some established threshold, we can apply to those accidental clicks a default discounting strategy by choosing among one of the discount factors computed for the other apps.

\subsection{Estimating non-accidental click rate}

To implement our proposed discounting strategy, we must accurately estimate the (binomial) proportion of non-ac\-ci\-den\-tal clicks $NACR$ of an app.
The most straightforward way is to use maximum likelihood estimate, na\-mely a single-point estimate $\widehat{NACR}_{\text{MLE}} \approx NACR$, which is the overall number of non-accidental ad clicks divided by the total number of ad clicks observed during a specific time window: {\small $\widehat{NACR}_{\text{MLE}} = \frac{\sum_{i\in \text{ads}}clicks_{i}^{\text{non-acc}}}{\sum_{i\in \text{ads}}clicks_i}$}.
This estimate however is not robust when we have an app with a low number of observations. 
To overcome this, we compute the confidence interval for $\widehat{NACR}_{\text{MLE}}$.

There exist several ways to compute a confidence interval for an estimate $\hat p$ of a binomial proportion $p$.\footnote{\small{In this setting, $p=NACR$ and $\hat p = \widehat{NACR}_{\text{MLE}}$.}} The normal approximation interval is the simplest and most common approach, and assumes the distribution of error of a binomially-distributed observation to be Gaussian.
This is computed as 
{\small $\hat p \pm z \sqrt{\frac{1}{n}\hat p(1-\hat p)}$}, where $\hat p$ is the proportion of successes in a Bernoulli trial process estimated from the statistical sample, $z$ is the $1 - \frac{1}{2}\alpha$ percentile of a standard Normal distribution, $\alpha$ is the error percentile and $n$ is the sample size. 

The normal approximation however does not always work. Several competing formulas are available that perform better, especially for situations with a small sample size and a proportion very close either to 0 or 1. 
The choice will depend on how important it is to use a simple and easy-to-explain interval versus the desire for better accuracy. 
As such, the  Agresti-Coull interval~\cite{BrownSS2001} is another approximate binomial confidence interval, which is more robust than the normal approximation interval. Given $X$ successes in $n$ Bernoulli trials, it defines the following quantities: {\small $\tilde{n} = n + z^2$} and {\small $\tilde{p} = \frac{1}{\tilde{n}}\left(X + \frac{1}{2}z^2\right)$}.
Then, a confidence interval for $p$ is given by: {\small $\tilde{p} \pm z \sqrt{\frac{1}{\tilde{n}}\tilde{p}\left(1 - \tilde{p} \right)}$}, where $z$ is the $1 - \frac{1}{2}\alpha $ percentile of a standard Normal distribution, as before.

\subsection{Comparing non-accidental click rates}

The proposed discounting strategy requires computing the ratio of non-accidental click rates between the app of interest and the pivoting app.
If the estimate of $NACR$ is not a single-point estimate such as $\widehat{NACR}_{\text{MLE}}$ but is a confidence interval, one way to compare two $NACR$ estimates is to take the ratio of their \emph{upper confidence bounds}:
\begin{equation}
\frac{NACR_{\textsf{app}}}{NACR_{\textsf{pivot}}} = \frac{\textsf{ucb}(NACR_{\textsf{app}})}{\textsf{ucb}(NACR_{\textsf{pivot}})} \label{eqn:ratiobound}
\end{equation}
where \textsf{ucb} is the upper confidence bound computed by the Agresti-Coull interval defined at the end of previous section.

We already stated that the pivoting app is assumed to be the one with the highest $NACR$.
However, for our discounting strategy to be robust it should also account for the case when an app is overperforming in terms of ``click value'' the pivoting app\footnote{\small{This may happen if the same pivoting app has been running for long and a new, better performing app slightly overtakes it.}}.
In such a case, we would like the discount factor $\frac{NACR_{\textsf{app}}}{NACR_{\textsf{pivot}}}$ to be greater than 1 only when we have a degree of confidence in it. 
One way to implement this is to require the overperforming app's {\em lower confidence bound} (\textsf{lcb}) being greater than the pivot's. 
We can therefore modify Equation~\ref{eqn:ratiobound} to: 
{\small \begin{equation}
\frac{NACR_{\textsf{app}}}{NACR_{\textsf{pivot}}} =  \text{max}\Bigg[   \frac{\textsf{lcb}(NACR_{\textsf{app}})}{\textsf{ucb}(NACR_{\textsf{pivot}})}, \text{min} \Bigg(  \frac{\textsf{ucb}(NACR_{\textsf{app}})} {\textsf{ucb}(NACR_{\textsf{pivot}}) } ,1  \Bigg)  \Bigg] \label{eqn:ratiobound2}
\end{equation}}
We make the following observations for Equation~\ref{eqn:ratiobound2}.
First, the ratio of non-accidental click rate will be greater than 1 only if the lower confidence bound of the app is greater than the upper confidence bound of the pivot.
Second, in case of a large sample with non-zero valid clicks the ratio of non-accidental click rate will converge to the ratio of single-point estimate (MLE). 
Third, in case of a small sample size the ratio of non-accidental click rate will be close to 1 indicating that we do not have enough data to suggest that the app of interest is any different from the pivoting app.
Fourth, there is still a minimum number of clicks needed for Equation~\ref{eqn:ratiobound2} to produce reliable results. 

When we have enough confidence that the ratio of non-accidental click rates between an app and the current pivot is greater than 1, we can update the pivot with that app and compute the discount factors using the new app as the new benchmark.

Next, we discuss through an example the impact on revenue of this smooth discounting strategy once implemented.  

\subsection{The impact of smooth discount factors}

We compute the discount factors for accidental ad clicks on the two datasets $\mathcal{D}_1$ and $\mathcal{D}_2$, described in Section~\ref{sec:discovery:data}. 
We consider all the ads impressed and clicked on all three apps, {\sf App 1}, {\sf App 2}, and {\sf App 3}.
To increase the confidence in our estimates, we discard ads with less than 40 clicks on \emph{each} app. 
We select the pivoting app as the one with the highest $NACR$, estimated either via MLE or with the Agresti-Coull estimator.
In both cases, {\sf App 1} is chosen.

\begin{table}[t]
\centering
\small
\begin{tabular}{cc|c|cl|}
\cline{3-4}
& & \multicolumn{2}{ c| }{{\bf Discount Factors}}\\
\cline{3-4}
& & \multicolumn{1}{ c| }{{MLE}} & \multicolumn{1}{ c| }{{Agresti-Coull}} \\ \cline{1-4}
\multicolumn{1}{ |c  }{\multirow{2}{*}{$\mathcal{D}_1$} } &
\multicolumn{1}{ |c| }{{\sf App 2}} & \multicolumn{1}{ c| }{0.72} & \multicolumn{1}{ c| }{0.79}  \\ \cline{2-4}
\multicolumn{1}{ |c  }{}                        &
\multicolumn{1}{ |c| }{{\sf App 3}} & \multicolumn{1}{ c| }{0.64} & \multicolumn{1}{ c| }{0.73}  \\ \cline{1-4}
\multicolumn{1}{ |c  }{\multirow{2}{*}{$\mathcal{D}_2$} } &
\multicolumn{1}{ |c| }{{\sf App 2}} & \multicolumn{1}{ c| }{0.66} & \multicolumn{1}{ c| }{0.75}  \\ \cline{2-4}
\multicolumn{1}{ |c  }{}                        &
\multicolumn{1}{ |c| }{{\sf App 3}} & \multicolumn{1}{ c| }{NA (not enough obs.)} & \multicolumn{1}{ c| }{NA (not enough obs.)}  \\ \cline{1-4}
\end{tabular}
\caption{\small{Discount factors computed using $t_{\text{median}}$, and MLE and Agresti-Coull estimates of $NACR$.}}
\label{tab:discounts}
\end{table}

Table~\ref{tab:discounts} shows the discount factors computed using $t_{\text{median}}$ as the dwell time thresholds of accidental clicks ($t_{\text{median}}=2.1$ seconds on $\mathcal{D}_1$; $t_{\text{median}}=2.2$ seconds on $\mathcal{D}_2$), and two different ratio of estimates of $NACR$, one obtained with MLE and the other using Agresti-Coull in combination with Equation~\ref{eqn:ratiobound2}.
Each row shows how much an advertiser should be charged for one accidental click on an ad shown on the app indicated by that row depending on the estimator used, providing that a valid (non-accidental) click on the same ad would cost 1. For example, looking at $\mathcal{D}_2$ if an ad click on {\sf App 2} originally costs $1\$$ to the advertiser, any accidental ad click on the same app will instead cost $0.75\$$ after discounting using the  Agresti-Coull estimate of $NACR$ from Equation~\ref{eqn:ratiobound2}. 

We observe that the discount factors are comparable across the two datasets when computed using the same strategy. 
Moreover, larger discounting happens when generated from a single-point MLE of non-accidental click rate; the discounting is smaller when computed using the Agresti-Coull confidence interval. 
Depending on how ag\-gres\-si\-ve  the discounting has to be, one or the other approach may be chosen.

At the beginning of this section, we discussed how with dataset $\mathcal{D}_3$ the potential revenue loss that would result from fully discarding accidental clicks when billing advertisers was too high. 
Now using Equation~\ref{eq:discountedcost}, the discount factor defined in Equation~\ref{eqn:ratiobound2}, and using {\sf App 1} as pivoting app, 
the revenue drop is reduced by about 73.1\%; allowing for advertisers to save money on likely less valuable clicks, while controlling for revenue impact for the ad network and publishers.
%
\section{Use case II: Filtering out accidental clicks when training an ad click model}
\label{sec:training}

Many ad networks improve the logic behind their ad serving algorithm through machine-learned models. At each ad request, these models provide a ranked list of ads to serve to maximise the overall \emph{expected revenue}, eCPM.\footnote{CPM stands for \emph{cost per mille (impressions)} and indicates the earnings gained every thousand ad impressions sold.} 
The eCPM is an estimate of the truly observed CPM, computed for each ad $i$ as $eCPM_i = cpc_i \times eCTR_i$. Each $cpc_i$ (or bid) is the price an advertiser is willing to pay for buying an impression, whereas $eCTR_i$ is the estimate of the click-through rate of ad $i$ ($CTR_i$). Estimating $eCPM_i$ means estimating $CTR_i$ for all ads $i$ in the ad network inventory.
Machine-learned models achieve exactly this task, \ie, they are trained on historical datasets of ad clicks to predict CTR from a feature-vector representations of serviceable ads.

Training models on datasets containing a ``large en\-ough'' ratio of accidental clicks may overestimate an ad CTR. This is because the estimated CTR becomes  ``inflated'' with accidental clicks, eventually leading to the selection of irrelevant ads to serve.
Filtering out accidental clicks when training machine-learned models may provide a more accurate selection of ads, resulting in higher revenue for the ad network and the publisher. 

Using our data-driven approach to identify accidental clicks, we compute a threshold 
for a large number of Yahoo mobile apps (Yahoo News, Yahoo Mail, etc). 
We compute per-app thresholds of accidental click, as we have a sufficiently large number of dwell time observations for each app. 

For each app, we then filter out clicks that are below the corresponding threshold for that app. The filtered clicks are not used to train the ad click model. We refer to this model as {\tt accidental\_click}. The model where no filtering of accidental clicks is our baseline mode, denoted as {\tt baseline}. 
We setup an online A/B testing experiment, where a fraction of Yahoo Gemini incoming ad traffic is split between a \emph{control} bucket and a \emph{variation} bucket. More specifically, the A/B test affects about $1.8\%$ of the overall ad serving traffic on the Yahoo apps considered. The traffic served by the control bucket is handled by the {\tt baseline} ad click model, whilst variation bucket dispatches traffic to the {\tt accidental\_click} model.
We thus compare the performance of the two models by measuring both CTR and CPM (click-through rate and revenue).

A significant lift is obtained using the \texttt{accidental\_cl\-ick} model of $+3.9\%$ compared to the \texttt{baseline} model (significance $\alpha = 0.01$ and $p$-value $\ll 0.01$ using one-tailed two-proportion z-test). 
This means that an ad click model trained on a cleaner dataset (\ie, without accidental clicks) leads to a better estimation of ad CTR. This trained click model is better at predicting ads that are more likely to be clicked, \ie, more relevant to users, as it is relies on ads that were clicked by users with an intent.
Similarly, we observe a statistically significant lift in CPM of $+0.2\%$ (level of significance $\alpha = 0.01$ and $p$-value $\ll 0.01$), which is partly due to the lift in CTR.\\

In this section, we showed that our data-driven approach for identifying accidental clicks is effective as a preprocessing step to training machine-learned ad click models, which ad networks leverage to estimate CTR to rank ads at serving time. 
%
\section{Related work}
\label{sec:relwork}

Various works investigated the role of dwell time on web pages.
Liu \ea~\cite{LiuSIGIR2010}, who modelled dwell time on web pages using a Weibull distribution~\cite{Papoulis2002},  
found that web browsing exhibits a significant ``negative aging'' phenomenon, suggesting that some initial screening has to be passed before a page is examined in detail by a user. 
They also demonstrated that dwell time distributions can be predicted purely using low-level web page features.
We extend this work -- focussing on ad (mobile) web pages -- by proposing a model of dwell time based on a mixture of distributions instead of Weibull. This allows us to capture three categories of ad clicks, \emph{accidental}, \emph{short}, and \emph{long}, where the focus of this paper is on accidental clicks.

Kim \ea~\cite{KimWSDM2014} presented a method to explain dwell time on search engine result pages.  
They estimate dwell time distributions for SAT (satisfied) or DSAT (dissatisfied) clicks for different click segments and use them to derive features to train a click-level satisfaction model.
Yi \ea~\cite{YiRecSys2014} use item-level dwell time as a proxy to quantify how likely a content item is relevant to a particular user in a recommender system.
Furthermore, Yin \ea~\cite{YinKDD2013} show how to enrich the user-vote matrix by converting dwell time on items into users' ``pseudo votes'' and then improve recommendation performance. 
All these works show that considering dwell time leads to improved decision-making tasks. In our work, our task is the identification of accidental clicks.

In the context of sponsored search \eg, \cite{BeckerCIKM2009,GrbovicSIGIR2016,RaghavanSIGIR2009,SculleyKDD2009,SodomkaWWW2013} and display advertising \eg, \cite{AzimiCIKM2012,BarajasCIKM2012,KaeLDMTA2011,RosalesWSDM2012}, studies have mostly focused on predicting ad performance in terms of CTR. An ad with a high ad CTR is considered to perform well, since it indicates that the ad attract users, who click on it.
CTR, however, does not account for the post-click experience, that is how users experience the ad landing page.
Dwell time on ad landing page has proven to be a good proxy of the post-click experience~\cite{BarbieriWWW2016,LalmasKDD2015}, reflecting the assumption that the longer the time user spends on the ad landing page the higher the chance he or she ``converts'' (\eg, by purchasing an item, registering to a mailing list, \etc), or simply the more likely is for the user to build an affinity with the brand~\cite{BeckerCIKM2009,RosalesWSDM2012}. These are cases of clicks that bring values to advertisers. 
Our work also looks at how dwell time relates to conversion rate, confirming what was shown in~\cite{Goldman2014GSP} that the former is a significant predictor of the latter, and hence a good proxy for measuring the ``value'' of a click.
We use dwell time as our proxy of the ad click value, and proposed a data-driven methodology to identify accidental clicks, \ie, ad clicks with very short dwell time that are not only valueless to advertisers, but also to machine-learning models trained to predict ad CTR.

Other studies have investigated click ``value'' in the context of online advertising, mostly to contrast with fraudulent activities perpetrated by dishonest advertisers and/or web publishers like \emph{click spam}~\cite{DaswaniHotBots2007,Stone-GrossIMC2011}.
To the best of our knowledge, this is the first work using dwell time to identify accidental clicks, in combination with a data-driven approach that can be applied to other domains, where it is important to quantify the value of a click.
%
\section{Conclusions}
\label{sec:conclusions}

In this paper, we propose a data-driven method to identify accidental clicks. An accidental click happens when a user that click on an ad, likely by mistake, is redirected to the ad landing page and bounce back without having seen the page. This type of clicks happens often on ads impressed on mobile devices.

We collect empirical dwell time observations from several Yahoo mobile apps for a large number of ads. 
We decompose the distribution of dwell time into a \emph{mixture of components}, with each component corresponding to a click category: \emph{accidental}, \emph{short}, and \emph{long}.
Representative statistics for the first component of each ad are then further aggregated to provide an overall estimate of the dwell time threshold of accidental clicks.

We assess the validity of our method when this is applied on two use cases.
First, we describe a technique that estimates a \emph{smooth} discounting factor, so that accidental clicks are not fully charged nor totally discarded at billing time. This allows for a trade-off between advertiser's satisfaction and potential revenue loss. 
Experiments conducted on different Yahoo mobile apps confirm that thresholds found are stable over time, and revenue loss can be mitigated by around 73.1\% using our discounting strategy compared to not charging at all accidental clicks.
Second, we demonstrate that an existing machine-learned ad click model used at serving time leads to better online performance if trained on datasets where accidental clicks are removed using our data-driven approach. We observe a positive and statistically significant lift on CTR and CPM (+3.9\% and +0.2\%, respectively) with the model trained without accidental clicks. 

As future work, we plan to look at the two other components of the mixture models, so as to estimate thresholds -- again in a data-driven manner -- for short and long clicks. The former are clicks that suggest that the user had a negative post-click experience, whereas the latter is an indication of a positive post-click experience. Being able to do so per-app would remove the often ad hoc setting of dwell time thresholds.
In addition to that, we would also like to investigate if the same (or similar) methodology proposed in this work could be used to assess the engagement of users with \emph{any} section of a web page or a mobile app (i.e., not only with ads shown). Having a rigorous, data-driven methodology to classify content on the basis of the time users spent on it (i.e., dwell time) might be useful to providers, who could in turn make better decision on which of their assets they should invest more.

\begin{acknowledgements} The authors would like to thank Michal Aharon and Marc Bron for their support in setting up the online A/B test, which allowed them to deploy and assess their approach on a second use case, \ie, the ad click model.
\end{acknowledgements}

\medskip
\emph{On behalf of all authors, the corresponding author states that there is no conflict of interest.}

\bibliographystyle{spbasic}

\end{document}